\documentclass[aps,prd,11pt,twocolumn,superscriptaddress,
            floatfix,nofootinbib,longbibliography,reprint]{revtex4-2}

\usepackage{graphicx}  
\graphicspath{{./figures/}}
\usepackage{bm}        
\usepackage{amssymb}   
\usepackage{amsmath}
\usepackage{mathrsfs}
\usepackage{relsize}
\usepackage[caption=false]{subfig}
\usepackage{mathtools}
\usepackage{xspace}
\usepackage{orcidlink}
\usepackage{soul}
\usepackage{enumitem}
\usepackage[final]{microtype}
\usepackage{multirow}
\usepackage{tabularx}
\usepackage{booktabs}
\usepackage{amsthm,amsmath,amssymb}

\usepackage{xcolor}
\usepackage{hyperref} 
\definecolor{dodgerblue}{HTML}{1E90FF}
\definecolor{DARK_ORANGE}{HTML}{FF5B00}
\definecolor{DARK_GREEN}{HTML}{02590F}
\definecolor{DARK_RED}{HTML}{B22222}
\definecolor{LIGHTER_DARK_RED}{HTML}{C86B6B}
\definecolor{dodgerblue}{HTML}{1E90FF}
\hypersetup{colorlinks=true, citecolor=DARK_RED,  urlcolor=dodgerblue, linkcolor=blue, pdfpagemode=FullScreen}

\newcommand{\orcidicon}[1]{\href{https://orcid.org/#1}{\includegraphics[height=\fontcharht\font`\B]{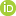}}}

\newcommand{\phXPHM}{\mbox{\textsc{IMRPhenomXPHM}}\xspace}

\newcommand{\bilby}{\textsc{Bilby}\xspace}

\newcommand{\dynesty}{\textsc{Dynesty}\xspace}

\begin{document}
\title{Gravitational Waves Population Study for Testing Kerr Hypothesis: are Non-Kerr Deformations Universal or Source-Dependent? }

\author{Debtroy~Das~\orcidlink{0009-0005-8097-7923}}
\affiliation{International Centre for Theoretical Physics Asia-Pacific (ICTP-AP), University of Chinese Academy of Sciences (UCAS), Beijing 100190, China.}
\affiliation{Taiji Laboratory for Gravitational Wave Universe (Beijing/Hangzhou), University of Chinese Academy of Sciences (UCAS), Beijing 100190, China.}

\author{Swarnim~Shashank~\orcidlink{0000-0003-3402-7212}}
\email{swarnim.shashank@astro.uni-tuebingen.de}
\affiliation{Institut f\"ur Astronomie und Astrophysik, Eberhard-Karls Universit\"at T\"ubingen, D-72076 T\"ubingen, Germany}
\affiliation{College of Fellows, Eberhard-Karls Universit\"at T\"ubingen, D-72070 T\"ubingen, Germany}

\author{Cosimo Bambi~\orcidlink{0000-0002-3180-9502}}
\email{bambi@fudan.edu.cn}
\affiliation{Center for Astronomy and Astrophysics, Department of Physics, Fudan University, Shanghai 200438, China}
\affiliation{Institute for Advanced Studies, New Uzbekistan University, Tashkent 100007, Uzbekistan}

\author{Yi-Ming Hu}
\affiliation{TianQin Research Center for Gravitational Physics and School of Physics and Astronomy, Sun Yat-sen University (Zhuhai Campus), Zhuhai 519082, China}

\date{\today}

\begin{abstract}
Gravitational-wave observations of binary black hole mergers can test the Kerr black hole hypothesis in the strong-field and dynamical regime. In our previous works, we constrained the deformation parameters $\alpha_{13}$ and $\epsilon_3$ of the Johannsen metric using individual inspiral-dominated events from GWTC-3 and GWTC-4 and found no evidence of deviations from the Kerr solution. Here we combine 17 of these events to test whether black holes are Kerr, share a common non-zero deformation, or have deformations that vary from source to source. We first estimate the Savage--Dickey Bayes factor of each event, and we test how it depends on the prior range and on the removal of single events. We then combine the event likelihoods assuming a universal deformation, yielding a tighter joint constraint. Next, we perform a simple hierarchical analysis by assuming that the event-level deformation parameters are drawn from a population with mean $\mu$ and intrinsic scatter $\sigma$. A universal deformation is constrained to $\alpha_{13}=-0.01^{+0.14}_{-0.12}$ and $\epsilon_3=0.06^{+0.59}_{-0.54}$. The means from the population are consistent with zero for both parameters, and the scatter is not resolved by the current data. Our results are consistent with the Kerr hypothesis, but they cannot yet distinguish a shared deformation from a source-dependent one.
\end{abstract}

\maketitle

\section{\label{sec:1}Introduction}

Einstein's theory of General Relativity (GR) was proposed at the end of 1915~\cite{Einstein:1916vd} and has so far successfully passed a large number of experimental tests without requiring any modification of its original form. For decades, the theory was mainly tested in the weak-field limit through experiments in the Solar System~\cite{Will:2014kxa}, while the strong-field regime remained almost completely unexplored. Over the last $\sim$10~years, however, we have begun to successfully test the predictions of GR in the strong-field regime as well.

Black holes are the sources of the strongest gravitational fields in the contemporary Universe and are therefore ideal laboratories for testing gravity in the strong-field regime~\cite{Bambi:2017khi,Bambi:2015kza,Bambi:2025kcr}. Until about $\sim$10~years ago, we had only indirect evidence for the existence of these objects. We knew of stellar-mass black holes in X-ray binaries and identified them as black holes simply because they were dark, compact objects with a mass exceeding the theoretical maximum limit for neutron stars~\cite{Rhoades:1974fn,Lattimer:2012nd}. We also knew of supermassive black holes in galactic nuclei and identified them as black holes because they were too massive, too compact, and too old to be clusters of neutron stars~\cite{Maoz:1997yd}. In the end, there was no evidence that these objects were the black holes predicted by GR, but there were certainly no alternative explanations within the framework of conventional physics.

Over the past $\sim$10~years, this picture has changed dramatically. Today we can probe the spacetime metric around black holes with X-ray observations~\cite{Bambi:2016sac,Cao:2017kdq,Tripathi:2018lhx,Tripathi:2020yts} and black hole imaging~\cite{EventHorizonTelescope:2020qrl,EventHorizonTelescope:2022xqj,Vagnozzi:2022moj}. We can simultaneously test the strong and dynamical regimes with gravitational-wave (GW) data from the coalescence of two black holes~\cite{LIGOScientific:2016lio,Yunes:2016jcc,LIGOScientific:2021sio}. Although current constraints on possible deviations from the predictions of GR may still appear weak compared with our ability to test atomic and particle physics, they have completely changed the paradigm and can be improved with the next generation of observational facilities.

The spacetime metric around astrophysical black holes is thought to be well approximated by the Kerr solution of GR~\cite{Kerr:1963ud} and to be completely characterized by only two parameters, namely the black hole mass and the black hole spin angular momentum. This is the result of the ``no-hair'' theorem, which is actually a family of theorems and holds under specific assumptions~\cite{Carter:1971zc,Robinson:1975bv,Chrusciel:2012jk}. The electric charge of an astrophysical black hole of a few solar masses or more is expected to be completely negligible, and any significant initial electric charge would be quickly almost neutralized by the highly ionized host environment~\cite{Bambi:2008hp}. The presence of nearby stars or of an accretion disk cannot appreciably change the spacetime metric around a black hole either~\cite{Bambi:2014koa}. Ultimately, the spacetime metric around an astrophysical black hole should be extremely close to the ideal Kerr solution, and macroscopic deviations from the Kerr background are possible only in the presence of new physics.

Generally speaking, if astrophysical black holes were not the Kerr black holes predicted by GR, we could have three possible scenarios of new physics~\cite{Bambi:2017khi,Herdeiro:2015waa}:
\begin{enumerate}
\item {\it Scenario 1: black holes have no hairs}. Astrophysical black holes are not the Kerr black holes of GR, but the no-hair theorem holds, so every object is still completely described by its mass and spin angular momentum. In this scenario, it is common for the black hole solution to include an extra parameter related to some new coupling of the underlying theory, which is therefore the same for every black hole.
\item {\it Scenario 2: black holes have hairs}. The no-hair theorem is violated, and different black holes can have different values of mass, spin angular momentum, and other parameters. In this scenario, deviations from the Kerr solution would not be universal but would depend on the individual object.
\item {\it Scenario 3: black holes have secondary hairs}. In this scenario, black holes have some charges, but their values are still determined by the black hole mass and/or spin, so some form of the no-hair theorem still holds. Black holes with the same mass and spin would have the same deviations from the Kerr geometry, but black holes with different masses and/or spins can appear to have different deformations from the Kerr solution.
\end{enumerate}

In Refs.~\cite{Das:2024mjq,Das:2026zyt}, we presented updated constraints on deviations from the Kerr metric using binary black hole (BBH) inspirals from the third and fourth Gravitational-Wave Transient Catalogs (GWTC-3 and GWTC-4). We employed a theory-agnostic framework in which non-Kerr effects of the Johannsen metric were incorporated as parameterized corrections to the GW phase within the post-Newtonian framework. We then inferred constraints on the deformation parameters $\alpha_{13}$ and $\epsilon_3$ from individual GW events in a selected sample. Our results are consistent with no deviation from the Kerr solution, thereby reinforcing the validity of Einstein's theory of GR and of the Kerr black hole hypothesis.

In the present work, we extend the analysis of Refs.~\cite{Das:2024mjq,Das:2026zyt} by combining different GW events into a joint analysis. First, under the assumption of a universal deformation parameter, we combine the event likelihoods to obtain a tighter joint constraint and a joint Bayes factor. Second, we perform a simple hierarchical analysis by assuming that the event-level deformation parameters are drawn from a population with mean $\mu$ and intrinsic scatter $\sigma$. In this picture, $\mu=\sigma=0$ corresponds to GR, $\sigma=0$ with $\mu \neq 0$ corresponds to a universal deformation (scenario~1), and $\sigma>0$ allows genuine source-to-source variation (scenario~2). A limitation of our current study is that the waveform assigns the same deformation parameter to both black holes in each binary. Therefore, the present analysis constrains an event-level effective deformation rather than separate hairs for the two component black holes. Since all GW events are from the coalescence of slow-rotating stellar-mass black holes, we do not try to test scenario~3 of black holes with secondary hairs.

The manuscript is organized as follows. In Section~\ref{sec:2}, we briefly review the key ingredients of our study: the non-Kerr metric employed in our work, the modified waveform during an inspiral phase, and the Bayesian inference used in our analysis. We derive observational constraints on the deformation parameters from individual events in Section~\ref{sec:3} and through a population study in Section~\ref{sec:4}. We discuss our results in Section~\ref{sec:5}.

\section{Mathematical background: Johanssen metric, waveform modifications and Bayesian inference}
\label{sec:2}

We describe possible departures from the Kerr geometry with the Johannsen metric~\cite{Johannsen:2013szh}. It is a phenomenological metric and represents a Kerr-like spacetime that is stationary, axisymmetric, asymptotically flat and regular outside the horizon, and it keeps a Carter-like constant of motion, so equations of motion are separable. Besides the mass and spin it carries free functions whose leading-order coefficients are $\alpha_{13}$, $\alpha_{22}$, $\alpha_{52}$ and $\epsilon_3$. Kerr is recovered when all of them vanish. As in our earlier work~\cite{Das:2024mjq,Das:2026zyt}, we vary $\alpha_{13}$ and $\epsilon_3$ one at a time and denote either of them by $\delta$. Requiring a regular spacetime with an event horizon bounds the deformation from below, described by equation (19) of our previous work~\cite{Das:2024mjq}.
 
A deformed spacetime changes the binding energy and orbital frequency of quasi-circular orbits, and therefore the rate at which an inspiral sweeps through frequency. We include this through the parametrized post-Einsteinian framework~\cite{Yunes:2009ke,Nissanke:2004er,Cardenas-Avendano:2019zxd}, writing the frequency-domain phase as 
\begin{equation}
\Psi_{\rm GW}(f) = \Psi^{\rm GR}_{\rm GW}(f) + \Psi^{\rm NK}_{\rm GW}(f).
\end{equation}

We keep the deformation's conservative effects up to 4PN order and assume that GW emission is as in GR~\cite{Carson:2020iik}. The additional phase due to the non-Kerr parameters are
\begin{eqnarray}
\Psi^{\rm NK}_{\rm GW}(\alpha_{13}) &=& -\frac{45\,\alpha_{13}\, u^{1/3}}{4\eta^{6/5}} + \frac{10\,\alpha_{13}\, u}{3\eta^{8/5}} - \frac{15\,\alpha_{13}}{4\eta^{4/5}u^{1/3}} \nonumber\\
&& - \frac{10\,\alpha_{13}\, u\log u}{3\eta^{8/5}}, \label{eq:psi_a13}\\
\Psi^{\rm NK}_{\rm GW}(\epsilon_{3}) &=& \frac{45\,\epsilon_{3}\, u^{1/3}}{8\eta^{6/5}} - \frac{5\,\epsilon_{3}\, u}{3\eta^{8/5}} + \frac{15\,\epsilon_{3}}{32\eta^{4/5}u^{1/3}} \nonumber\\
&& + \frac{5\,\epsilon_{3}\, u\log u}{3\eta^{8/5}}, \label{eq:psi_e3}
\end{eqnarray}
with $\eta$ the symmetric mass ratio, $u=\eta^{3/5}\pi m f$, $m$ the total mass and $f$ the GW frequency.
 
Two features of Eqs.~\eqref{eq:psi_a13}--\eqref{eq:psi_e3} are important to be noted. First, the two parameters enter at the same PN orders, with coefficients in the ratio $-2$ for three of the four terms and $-8$ for the $u^{-1/3}$ term. A deformation in one can therefore be nearly mimicked by the other, and fitting both at once yields an elongated, essentially unconstrained posterior~\cite{Das:2024mjq}. This is why we analyse them separately. Second, we assign the same deformation parameter $\delta$ to both black holes
in a binary, where $\delta$ stands for either $\alpha_{13}$ or $\epsilon_3$, so each event constrains an effective, event-level deformation rather than two independent ones.

The non-Kerr waveform model is built by modifying a standard GR template. We use the frequency-domain \phXPHM model~\cite{Pratten:2020ceb} for the GR signal $\tilde h_{\rm GR}(f;\boldsymbol{\theta})$. A custom frequency-domain source model is generated by multiplying $\tilde h_{\rm GR}(f;\boldsymbol{\theta})$ by the phase factor $e^{i\Psi^{\rm NK}_{\rm GW}}$ as, 
\begin{equation}
    \tilde h(f;\boldsymbol{\theta},\delta) = \tilde h_{\rm GR}(f;\boldsymbol{\theta})\, e^{\,i\,\Psi^{\rm NK}_{\rm GW}(f;\eta,m,\delta)} .
\end{equation}
The amplitude is kept at 0PN and unmodified, since the ppE framework used here accounts only for phase modifications.

The ppE correction is valid only in the inspiral, so only the inspiral part of the data is analyzed. The deformation parameter is sampled together with the standard BBH parameters (masses, spins, distance, sky location, orientation, etc.) using Bayesian inference, performed with \bilby~\cite{Ashton:2018jfp} using the \dynesty nested-sampling algorithm~\cite{Speagle:2019ivv}. We use LVK-consistent priors for the BBH parameters~\cite{Romero-Shaw:2020owr} and a uniform prior on $\delta$. The event-dependent lower bounds are imposed through a conversion function in \bilby, computed from the sampled component masses and spins at each step. We use the likelihood function,
\begin{equation}
    \mathcal{L}(d\,|\,\boldsymbol{\theta}) \propto \exp\!\left(-\sum_k \frac{2\,|d_k - h_k(\boldsymbol{\theta})|^2}{T S_k}\right),
\end{equation}
where $d$ is the GW data, $h(\boldsymbol{\theta})$ denotes the gravitational waveform for the BBH signal with parameters $\boldsymbol{\theta}$, $T$ is the segment duration and $S_k$ the noise power spectral density and $k$ signifies the frequency dependence of the data, waveform, and PSD. The posterior is $p(\boldsymbol{\theta}\,|\,d)\propto \mathcal{L}(d\,|\,\boldsymbol{\theta})\,\pi(\boldsymbol{\theta})$. The resulting single-event priors and posteriors are the inputs to everything that follows.

\section{Preliminary analysis of the single-event posteriors}
\label{sec:3}

We start from the 17 events already analyzed in our earlier work: six from GWTC-3~\cite{Das:2024mjq} and eleven from the new events from  GWTC-4~\cite{Das:2026zyt}. They were chosen because their inspirals are well observed and well described by an aligned-spin, quasi-circular PN picture (detector-frame total mass below $80\,M_\odot$, network SNR above 10, spin-precession parameter $\chi_{\rm p}<0.5$). In every case, $\alpha_{13}$ and $\epsilon_3$, varied one at a time, are consistent with zero, and the GWTC-4 events tighten the bounds considerably.
 
Consistency with zero in each event, however, is a weaker statement than it sounds. It only says that any deformation is smaller than that event's resolution, which is of order unity for $\alpha_{13}$ and a few for $\epsilon_3$. Three very different pictures all predict the same single-event outcome: (i)~every black hole is Kerr; (ii)~all black holes share one small non-zero deformation, as would follow from a generalized no-hair theorem; (iii)~the deformation varies from source to source, as expected if black holes carry additional hair. For cases (ii) and (iii), telling these pictures apart requires combining events in a way that does not presuppose the answer, which is the motivation for the population-level treatment of Sec.~\ref{sec:4}~\cite{Zimmerman:2019wzo,Isi:2019asy,Isi:2022cii}.
 
Before doing so, we ask what the single-event results already say and how much we can trust them as building blocks. We analyze the output of each event to obtain a Bayes factor between Kerr and the deformed model, assess how these numbers depend on the range of deformations we allow \emph{a priori}, and test whether any single event dominates the combination.

\subsection{How strongly does each event favour Kerr?}
\label{sec:3A}

Kerr is the special case $\delta=0$ of the Johannsen metric, so comparing them is a question of Occam's razor~\cite{jefferys1992ockham, mackay2003information}. The deformed model can fit the data at least as well as Kerr, but it must spread its predictions over the entire range of allowed deformations beforehand and is penalized for doing so. The Bayes factor $\mathrm{BF}_{01}$ measures this trade-off. Values above one favor Kerr, values below one favor a deformation.

Because Kerr is nested inside the deformed model, $\mathrm{BF}_{01}$ can be read directly from the existing posterior samples, with no new inference runs. It is the ratio of the posterior probability density at $\delta=0$ to the prior density at that point. This is called the Savage--Dickey density ratio~\cite{dickey1971weighted, verdinelli1995computing, wagenmakers2010bayesian}, defined as 
\begin{equation}
    \text{BF}_{01}^{(i)} = \frac{p_i(\delta = 0 \mid d_i, \mathcal{M}_1)}{\pi_i(\delta = 0 \mid \mathcal{M}_1)}.
\end{equation}

Here, $\pi_i(\theta \mid \mathcal{M}_1)$ represents the single-event prior distribution assigned to the Johannsen deformation parameter, and $p_i(\theta \mid d_i, \mathcal{M}_1)$ is the marginal posterior density reconstructed from the single-event posterior samples. We estimate the posterior density at zero with a kernel density estimate of the samples~\cite{Silverman1986}. For uniform priors bounded by $[\delta_{\text{min}}, \delta_{\text{max}}]$, the prior density at the reference point is constant, given simply by $\pi_i(0) = 1 / (\delta_{\text{max}} - \delta_{\text{min}})$.

Table~\ref{tab:bfs} summarizes the prior boundaries $[\delta_{\text{min}}, \delta_{\text{max}}]$ and event-level Bayes factors $\text{BF}_{01}$, for both analyses where $\alpha_{13}$ and $\epsilon_3$ are individually constrained while the other is set to 0. In Ref.~\cite{Das:2024mjq}, which used GWTC-3 data, we adopted a uniform prior on $\delta$ over the range $[-5,5]$ for all events and for both deformation
parameters. Values beyond this interval would correspond to large deviations from GR, which are not expected. For the GWTC-4 events of Ref.~\cite{Das:2026zyt}, we anticipated tighter constraints due to the improved detector sensitivity. Therefore, we started each analysis with a narrower prior range, which reduces the computational cost. Whenever the posterior of $\delta$
reached a prior boundary, we repeated the analysis with a wider range within the interval $[-5,5]$.

\begin{table}[h]
\caption{\label{tab:bfs} Single-event Savage--Dickey Bayes factors ($\text{BF}_{01}$) favoring GR over non-GR deformations, along with individual static prior boundaries across events from O1 to O4.}
\begin{tabular}{lcc}
\hline\hline
\textbf{Event Name} & \textbf{Prior Boundaries} & \textbf{Savage--Dickey $\text{BF}_{01}$} \\ \hline
\multicolumn{3}{c}{\textit{Section 1: $\alpha_{13}$ Free, $\epsilon_3 = 0$}} \\ \hline
GW150914 & $[-5.0, 5.0]$ & 10.147 \\
GW170814 & $[-5.0, 5.0]$ & 12.837 \\
GW190630\_185205 & $[-5.0, 5.0]$ & 8.930 \\
GW190828\_063405 & $[-5.0, 5.0]$ & 5.473 \\
GW200112\_155838 & $[-5.0, 5.0]$ & 12.822 \\
GW200311\_115853 & $[-5.0, 5.0]$ & 4.988 \\
GW230605\_065343 & $[-2.0, 2.0]$ & 0.333 \\
GW230627\_015337 & $[-1.0, 1.0]$ & 7.303 \\
GW230628\_231200 & $[-5.0, 5.0]$ & 11.090 \\
GW230814\_230901 & $[-5.0, 5.0]$ & 31.313 \\
GW230904\_051013 & $[-2.0, 2.0]$ & 0.594 \\
GW230927\_153832 & $[-3.0, 3.0]$ & 10.863 \\
GW231020\_142947 & $[-1.5, 1.5]$ & 2.719 \\
GW231108\_125142 & $[-2.0, 2.0]$ & 2.348 \\
GW231226\_101520 & $[-5.0, 5.0]$ & 12.715 \\
GW231231\_154016 & $[-1.5, 1.5]$ & 2.404 \\
GW240109\_050431 & $[-3.0, 2.0]$ & 3.575 \\ \hline
\multicolumn{3}{c}{\textit{Section 2: $\epsilon_3$ Free, $\alpha_{13} = 0$}} \\ \hline
GW150914 & $[-5.0, 5.0]$ & 2.329 \\
GW170814 & $[-5.0, 5.0]$ & 3.033 \\
GW190630\_185205 & $[-5.0, 5.0]$ & 2.092 \\
GW190828\_063405 & $[-5.0, 5.0]$ & 1.318 \\
GW200112\_155838 & $[-5.0, 5.0]$ & 3.085 \\
GW200311\_115853 & $[-5.0, 5.0]$ & 1.327 \\
GW230605\_065343 & $[-3.0, 3.0]$ & 0.105 \\
GW230627\_015337 & $[-1.5, 1.2]$ & 2.352 \\
GW230628\_231200 & $[-5.0, 3.0]$ & 0.822 \\
GW230814\_230901 & $[-5.0, 5.0]$ & 7.149 \\
GW230904\_051013 & $[-5.0, 4.5]$ & 0.299 \\
GW230927\_153832 & $[-3.0, 3.0]$ & 2.654 \\
GW231020\_142947 & $[-5.0, 3.5]$ & 1.768 \\
GW231108\_125142 & $[-5.0, 3.5]$ & 1.228 \\
GW231226\_101520 & $[-5.0, 5.0]$ & 3.136 \\
GW231231\_154016 & $[-3.0, 5.0]$ & 1.547 \\
GW240109\_050431 & $[-5.0, 5.0]$ & 1.772 \\ \hline\hline
\end{tabular}
\end{table}

A useful way to interpret the Savage--Dickey Bayes factors is that if the posterior is centered close to zero, $\mathrm{BF}_{01} = 10$ roughly means the data have ruled out about nine-tenths of the territory the deformed model initially explored. Table~\ref{tab:bfs} shows that nearly every event favors Kerr. For $\alpha_{13}$, 15 of the 17 events have $\mathrm{BF}_{01}>1$, mostly between 5 and 13, with the loudest event, GW230814\_230901, reaching 31. For $\epsilon_3$, 14 events favor Kerr, typically by factors of only 1 to 3, with GW230814\_230901 again the best at 7. If Kerr and the deformed model are considered equally likely beforehand, $\mathrm{BF}_{01}$ is directly the odds in favor of Kerr. A factor of 10 means Kerr is ten times more probable than the non-Kerr model given that event alone. The difference in the inferred values of $\text{BF}_{01}^{(i)}$ between the two parameters reflects how strongly each leaves its imprint on the phase. The PN coefficients of the $\alpha_{13}$ in the GW phase terms are about twice as large as the PN coefficients of $\epsilon_3$ terms (Eqs.~\eqref{eq:psi_a13}--\eqref{eq:psi_e3}), so the same data constrain it more tightly.

\begin{table*}[ht]
\caption{\label{tab:prior_sensitivity_alpha13} Prior predictive sensitivity analysis for $\alpha_{13}$ (with $\epsilon_3 = 0$), showing Bayes factors ($\text{BF}_{01}$) across varying prior widths $W \in \{1.0, 2.0, 5.0, 10.0, 20.0\}$.}
\setlength{\tabcolsep}{5mm}
\begin{tabular}{lccccc}
\hline\hline
\textbf{Event Name} & \textbf{$W = 1.0$} & \textbf{$W = 2.0$} & \textbf{$W = 5.0$} & \textbf{$W = 10.0$} & \textbf{$W = 20.0$} \\ \hline
GW150914 & 1.013 & 2.025 & 5.063 & 10.126 & 20.253 \\
GW170814 & 1.281 & 2.562 & 6.406 & 12.812 & 25.623 \\
GW190630\_185205 & 0.891 & 1.782 & 4.456 & 8.912 & 17.825 \\
GW190828\_063405 & 0.546 & 1.092 & 2.731 & 5.462 & 10.924 \\
GW200112\_155838 & 1.280 & 2.559 & 6.398 & 12.796 & 25.592 \\
GW200311\_115853 & 0.498 & 0.996 & 2.489 & 4.978 & 9.955 \\
GW230605\_065343 & 0.083 & 0.166 & 0.415 & 0.829 & 1.658 \\
GW230627\_015337 & 3.651 & 7.303 & 18.257 & 36.514 & 73.027 \\
GW230628\_231200 & 1.107 & 2.214 & 5.534 & 11.068 & 22.136 \\
GW230814\_230901 & 3.125 & 6.250 & 15.625 & 31.251 & 62.501 \\
GW230904\_051013 & 0.154 & 0.309 & 0.772 & 1.545 & 3.090 \\
GW230927\_153832 & 1.811 & 3.621 & 9.053 & 18.106 & 36.211 \\
GW231020\_142947 & 0.935 & 1.871 & 4.676 & 9.353 & 18.705 \\
GW231108\_125142 & 0.587 & 1.174 & 2.935 & 5.870 & 11.741 \\
GW231226\_101520 & 1.269 & 2.538 & 6.345 & 12.690 & 25.380 \\
GW231231\_154016 & 0.801 & 1.603 & 4.007 & 8.014 & 16.027 \\
GW240109\_050431 & 0.715 & 1.430 & 3.575 & 7.150 & 14.300 \\ \hline\hline
\end{tabular}
\end{table*}

\begin{table*}[ht]
\centering
\caption{Prior predictive sensitivity analysis for $\epsilon_3$ (with $\alpha_{13} = 0$), showing Bayes factors ($\text{BF}_{01}$) across varying prior widths $W \in \{1.0, 2.0, 5.0, 10.0, 20.0\}$.}
\label{tab:prior_sensitivity_epsilon3}
\setlength{\tabcolsep}{5mm}
\begin{tabular}{lccccc}
\hline\hline
\textbf{Event Name} & \textbf{$W = 1.0$} & \textbf{$W = 2.0$} & \textbf{$W = 5.0$} & \textbf{$W = 10.0$} & \textbf{$W = 20.0$} \\ \hline
GW150914 & 0.232 & 0.465 & 1.162 & 2.325 & 4.649 \\
GW170814 & 0.303 & 0.605 & 1.514 & 3.027 & 6.054 \\
GW190630\_185205 & 0.209 & 0.418 & 1.044 & 2.088 & 4.176 \\
GW190828\_063405 & 0.132 & 0.263 & 0.658 & 1.316 & 2.632 \\
GW200112\_155838 & 0.308 & 0.616 & 1.539 & 3.079 & 6.158 \\
GW200311\_115853 & 0.132 & 0.265 & 0.662 & 1.324 & 2.648 \\
GW230605\_065343 & 0.017 & 0.035 & 0.087 & 0.175 & 0.349 \\
GW230627\_015337 & 0.871 & 1.742 & 4.355 & 8.710 & 17.420 \\
GW230628\_231200 & 0.103 & 0.206 & 0.514 & 1.028 & 2.056 \\
GW230814\_230901 & 0.715 & 1.430 & 3.574 & 7.149 & 14.297 \\
GW230904\_051013 & 0.031 & 0.063 & 0.157 & 0.315 & 0.630 \\
GW230927\_153832 & 0.442 & 0.885 & 2.212 & 4.424 & 8.847 \\
GW231020\_142947 & 0.214 & 0.429 & 1.072 & 2.144 & 4.287 \\
GW231108\_125142 & 0.144 & 0.289 & 0.722 & 1.444 & 2.889 \\
GW231226\_101520 & 0.314 & 0.627 & 1.568 & 3.136 & 6.273 \\
GW231231\_154016 & 0.193 & 0.387 & 0.967 & 1.934 & 3.867 \\
GW240109\_050431 & 0.177 & 0.354 & 0.886 & 1.772 & 3.544 \\ \hline\hline
\end{tabular}
\end{table*}

Only a few events lean the other way. GW230605\_065343 and GW230904\_051013 have $\mathrm{BF}_{01}<1$ for both parameters, and GW230628\_231200 does so for $\epsilon_3$. These are the events whose posteriors peak away from zero. We discuss possible reasons in Sec.~\ref{sec:3C}.

Multiplying the single-event factors gives $\sim 10^{12}$ for $\alpha_{13}$ and $\sim 2\times10^{3}$ for $\epsilon_3$. The product of single-event Bayes factors is the Bayes factor for a model in which every event has its own independent deformation drawn from the same broad prior~\cite{Zimmerman:2019wzo,Isi:2022cii}. Therefore, it measures how decisively the data reject unrestricted source-to-source variation across the full prior range, rather than a single shared deformation, which requires combining the likelihoods themselves and is treated in Sec.~\ref{sec:4}.

\subsection{Dependence on the allowed range of deformations}
\label{sec:3B}

A Bayes factor depends on how much freedom we give the deformed model. Suppose we allow $\delta$ to take any value between $-10$ and $+10$, and the data show no sign of a deformation. The deformed model is then heavily penalized because most of what it allowed was never observed. If we had allowed only $\pm0.5$, the same data would have penalized it much less. In the analysis of Ref.~\cite{Das:2024mjq} and Ref.~\cite{Das:2026zyt}, the posteriors are narrow compared to the prior, so the Bayes factor grows in proportion to the prior width $W$. To demonstrate this, we recompute each event's Bayes factor for $W$ over the range 1 to 20. This also puts all events on the same footing, which Table~\ref{tab:bfs} does not, since the event-specific ranges there differ. Table~\ref{tab:prior_sensitivity_alpha13} and Table~\ref{tab:prior_sensitivity_epsilon3} summarize the prior-predictive test results for the free $\alpha_{13}$ and free $\epsilon_3$ models, respectively. 

The linear growth with $W$ is the Occam penalty~\cite{gabry2019visualization,jefferys1992ockham,mackay2003information,murray2005note} at work and should not be read as growing evidence for Kerr. A simple way to read the tables is through the \emph{break-even width}, the prior width $W$ at which an event stops preferring Kerr, i.e., where $\mathrm{BF}_{01}\leq1$. It tells us how large a deformation the event cannot rule out. For $\alpha_{13}$, most events already have $\mathrm{BF}_{01}\approx1$ at $W=1$. So a single event distinguishing Kerr from a deformation with a distribution of size $|\alpha_{13}|\lesssim0.5$ is difficult. Only the two best events, GW230627\_015337 and GW230814\_230901, favor Kerr strongly at this width. For $\epsilon_3$, the constraints are weaker. At $W=1$, no event favors Kerr, and a typical event needs $W\sim3$ to 8 before it does. GW230605\_065343 and GW230904\_051013 are the exceptions. They prefer a deformation unless the prior is made very wide. Their break-even widths are about 12 and 6 for $\alpha_{13}$, and  $\sim60$ and $\sim30$ for $\epsilon_3$.

These widths do not depend on which prior one chooses, so they are the most useful numbers in this test. They also change how we should read the stacked Bayes factors of Sec.~\ref{sec:3A}. Those numbers indicate how well the data can resolve deformation ~\cite{Isi:2022cii}. They do not give the probability that Kerr is correct.

\subsection{Does any single event dominate?}
\label{sec:3C}

When many events are combined, a few loud ones can end up dictating the answer. To check, we remove each event in turn and recompute the stacked Bayes factor. This analysis is called a leave-one-out (LOO) test~\cite{stone1974cross,vehtari2017practical,vehtari2026loo}. The LOO population analysis proceeds by calculating the global Bayes factor $\text{BF}_{01}^{(\text{full})}$ as the product of the event-wise Savage-Dickey Bayes factors and then omitting one event $d_k$ at a time from the full set of $N$ events. For each sub-catalog of $N-1$ events, we re-evaluate the joint Bayes factor $\text{BF}_{01}^{(\setminus k)}$ and compute the percentage shift relative to the global catalog baseline $\text{BF}_{01}^{(\text{full})}$, 

\begin{equation}
    \text{Shift}\ (\%) = \left( \frac{\text{BF}_{01}^{(\setminus k)} - \text{BF}_{01}^{(\text{full})}}{\text{BF}_{01}^{(\text{full})}} \right) \times 100.
\end{equation}

A severe negative percentage shift indicates that event $d_k$ was heavily contributing to the global evidence in favor of GR. In contrast, a positive shift implies that removing event $d_k$ actually increases the joint preference for GR, i.e., the omitted event was mildly dragging down the global Bayes factor. Because the stacked factor is a product, removing an event just divides it by that event's own factor, so the test is equivalent to asking how many orders of magnitude each event contributes.

\begin{table*}[ht]
\caption{\label{tab:loo} Leave-one-out population sensitivity analysis for the non-GR deformation parameters $\alpha_{13}$ (with global $\text{BF}_{01}^{(\text{full})} = 9.90 \times 10^{11}$) and $\epsilon_3$ (with global $\text{BF}_{01}^{(\text{full})} = 1711.50$). For each omitted event, the table reports the estimated signal-to-noise ratio (SNR), the recalculated population Bayes factor ($\text{BF}_{01}^{(\setminus k)}$), and the corresponding percentage shift relative to the global population value.}
\setlength{\tabcolsep}{4mm}
\begin{tabular}{l ccc ccc}
\hline \hline
 & \multicolumn{3}{c}{\textbf{Section 1: Free $\alpha_{13}$ ($\epsilon_3 = 0$)}} & \multicolumn{3}{c}{\textbf{Section 2: Free $\epsilon_3$ ($\alpha_{13} = 0$)}} \\
\cline{2-4} \cline{5-7}
\textbf{Omitted Event $k$} & \textbf{Est. SNR} & \textbf{$\text{BF}_{01}^{(\setminus k)}$} & \textbf{\% Shift} & \textbf{Est. SNR} & \textbf{$\text{BF}_{01}^{(\setminus k)}$} & \textbf{\% Shift} \\ \hline

GW150914 & 24.478 & $9.75 \times 10^{10}$ & $-90.1$ & 24.475 & 734.789 & $-57.1$ \\
GW170814 & 17.227 & $7.71 \times 10^{10}$ & $-92.2$ & 17.254 & 564.276 & $-67.0$ \\
GW190630\_185205 & 15.410 & $1.11 \times 10^{11}$ & $-88.8$ & 15.367 & 818.100 & $-52.2$ \\
GW190828\_063405 & 16.487 & $1.81 \times 10^{11}$ & $-81.7$ & 16.532 & 1298.071 & $-24.2$ \\
GW200112\_155838 & 18.669 & $7.72 \times 10^{10}$ & $-92.2$ & 18.630 & 554.784 & $-67.6$ \\
GW200311\_115853 & 16.508 & $1.98 \times 10^{11}$ & $-80.0$ & 16.472 & 1290.230 & $-24.6$ \\
GW230605\_065343 & 29.650 & $2.97 \times 10^{12}$ & $+200.0$ & 29.615 & 16344.646 & $+855.0$ \\
GW230627\_015337 & 29.727 & $1.36 \times 10^{11}$ & $-86.3$ & 29.739 & 727.779 & $-57.5$ \\
GW230628\_231200 & 25.552 & $8.92 \times 10^{10}$ & $-91.0$ & 18.878 & 2081.011 & $+21.6$ \\
GW230814\_230901 & 40.123 & $3.16 \times 10^{10}$ & $-96.8$ & 40.132 & 239.413 & $-86.0$ \\
GW230904\_051013 & 7.519 & $2.51 \times 10^{12}$ & $+153.9$ & 7.532 & 5720.833 & $+234.3$ \\
GW230927\_153832 & 19.660 & $9.11 \times 10^{10}$ & $-90.8$ & 19.643 & 644.837 & $-62.3$ \\
GW231020\_142947 & 11.588 & $3.78 \times 10^{11}$ & $-61.8$ & 11.559 & 967.787 & $-43.5$ \\
GW231108\_125142 & 11.804 & $4.21 \times 10^{11}$ & $-57.4$ & 11.817 & 1393.988 & $-18.6$ \\
GW231226\_101520 & 34.741 & $7.78 \times 10^{10}$ & $-92.1$ & 34.736 & 545.699 & $-68.1$ \\
GW231231\_154016 & 13.538 & $4.12 \times 10^{11}$ & $-58.4$ & 13.528 & 1106.412 & $-35.4$ \\
GW240109\_050431 & 10.364 & $2.77 \times 10^{11}$ & $-72.0$ & 10.365 & 965.745 & $-43.6$ \\ \hline\hline
\end{tabular}
\end{table*}

Table~\ref{tab:loo} summarizes the LOO sensitivity test for both the free $\alpha_{13}$ and free $\epsilon_3$ model configurations. For $\alpha_{13}$, the stacked Bayes factor totals about 12 orders of magnitude. The largest single contribution comes from GW230814\_230901 (SNR $\approx40$). For $\epsilon_3$, the total is about 3 orders, and the same event supplies the strongest contribution. Without it, the other 16 events still give $3\times10^{10}$ and $\approx240$.

The biggest contributors are not simply the loudest events. GW170814 and GW200112\_155838 (SNR 17--19) add as much to $\alpha_{13}$ as GW231226\_101520 (SNR 35). For a test based on the inspiral phase, the number of inspiral cycles seems to matter as much as the total SNR.

Removal \emph{raises} the $\text{BF}_{01}^{(\setminus k)}$ for two events, GW230605\_065343 and GW230904\_051013. They are the only events that individually lean towards a deformation, so they pull the stacked factor down. GW230628\_231200 shows a milder version of this in the $\epsilon_3$ case. GW230605\_065343 is the more interesting of the three. The $\alpha_{13}$ posterior for this event is narrow and has an inferred value of $-0.30$, with the 99\% interval just reaching zero. Its $\epsilon_3$ posterior is offset the other way, $+1.25$, with zero again near the 99\% interval edge~\cite{Das:2026zyt}. Because the phase corrections for both the deformation parameters are nearly degenerate (Eqs.~\eqref{eq:psi_a13}--\eqref{eq:psi_e3}), opposite signs mean the same phase shift.

We know of no dedicated study of either GW230605\_065343 or GW230904\_051013. The LVK parametrized tests of GWTC-4.0 include both and report nothing unusual for either~\cite{LIGOScientific:2026fcf}. They do single out GW230628\_231200. For some inspiral parameters, the FTI analysis places GR outside the 90\% credible interval (it remains within the 92.7\% credible interval). For some post-inspiral parameters, TIGER does the same (GR is inside 94.3\%), although its Bayes factors still favor GR. The LVK regard such deviations as expected by chance among the 91 events tested~\cite{LIGOScientific:2026fcf}. Our mild preference for non-zero $\epsilon_3$ in this event fits that picture.

Ref.~\cite{LIGOScientific:2026fcf} lists several ways to move a deviation posterior away from zero without new physics. These include low-SNR noise, degeneracies with source parameters, waveform modeling, PN-truncation systematics, and astrophysical assumptions. The leading term of our deformation is at 2PN relative order (the $u^{-1/3}$ term in Eqs.~\eqref{eq:psi_a13}--\eqref{eq:psi_e3}), where the GR phase already depends on the mass ratio. A mis-measured mass ratio or spin can therefore partly mimic a deformation. Our injection study in Ref.~\cite{Das:2026zyt} shows mild correlations between $\delta$ and the chirp mass. GW230904\_051013 has the lowest chirp mass of the three and also the very low SNR. However, for GW230605\_065343 (SNR $\approx 30$), we are unsure what is causing the posteriors of the deformation parameters to peak away from zero.

A finite catalog adds its own scatter. Among 17 events, the chance that at least one has a 99\% interval barely touching zero is about one in five. Ignoring this can make GR look excluded, or confirmed, with false confidence~\cite{Pacilio:2023uef}.

To summarize, these tests show that no single event dictates the stacked evidence, and that the single-event results are suitable inputs for a population analysis. The few events that favor a deformation do so only mildly. Effects, such as noise or waveform systematics, may explain them, but we have not identified the cause for any of them. We now use these results to
ask whether the data prefer no deformation, a shared one, or one that varies across sources.

\section{Population analysis}
\label{sec:4}

The single-event posteriors are consistent with Kerr, but they cannot tell us which of the scenarios of Sec.~\ref{sec:1} is realized. For that, we must
combine events. There are two natural ways to do so, and each assumes how $\delta$ behaves across black holes~\cite{Vitale:2020aaz,Smith:2020lkj}. These approaches correspond to Scenarios 1 and 2. In Scenario 1, a new coupling of the underlying theory is the same for every black hole, so all events share one value of $\delta$. We can then multiply their likelihoods and the events add up and the bound tightens. In Scenario 2, $\delta$ depends on the individual object, so pooling is not appropriate. Rather, we inspect how $\delta$ is distributed over the population. In the hierarchical model, Kerr is $\mu=\sigma=0$, Scenario 1 is $\sigma=0$ with $\mu\neq0$, and Scenario 2 is $\sigma>0$. We do not test Scenario 3, since our events are slowly rotating stellar-mass black holes. A $\delta$ that depends on mass or spin would also
show up as scatter. We discuss this point in Sec.~\ref{sec:5}.

We first treat the universal case in Sec.~\ref{sec:4A} and then the source-to-source variation case in Sec.~\ref{sec:4B}. Both reuse the existing single-event posterior samples, so no new parameter-estimation runs are needed.

\subsection{Universal deformation: combining event likelihoods}
\label{sec:4A}

\begin{figure}[h]
    \centering
	\includegraphics[width=\linewidth]{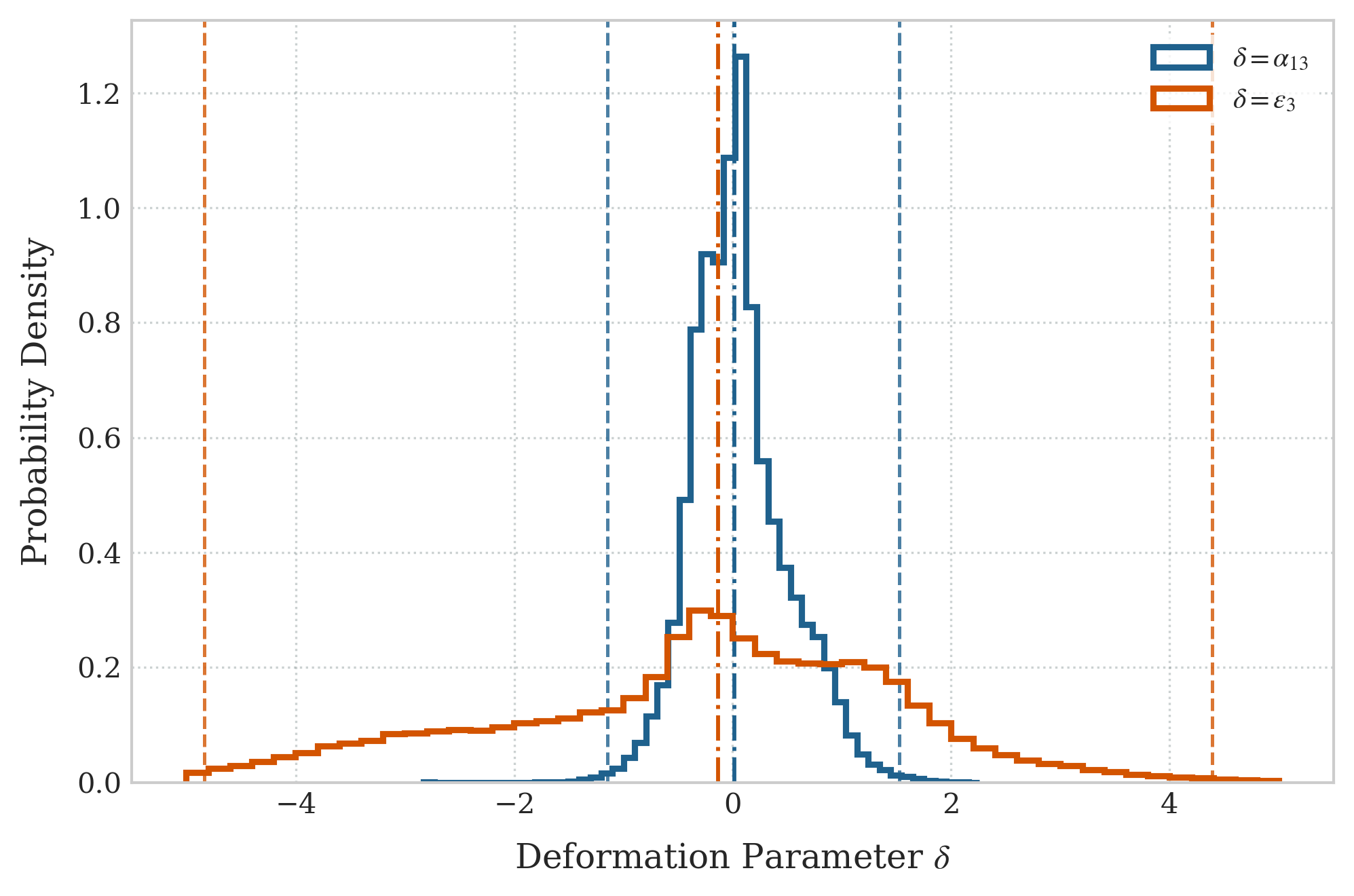} 
    \caption{Linear stacking of the single-event posteriors of the 17 events. Samples from all events are pooled with equal weight, without correction for differing prior ranges. The blue histogram shows $\alpha_{13}$, and the orange histogram shows $\epsilon_3$. Dot-dashed lines mark the medians and dashed lines mark the $99.7\%$ intervals, in the corresponding colors.}
    \label{fig:direct}
\end{figure}

If all black holes share the same $\delta$, we can combine the events into a single constraint. For $N$ independent events,
\begin{equation}
p(\delta\,|\,d_1,\dots,d_N)\;\propto\;\pi(\delta)\prod_{i=1}^{N}\mathcal{L}_i(d_i\,|\,\delta),
\end{equation}
where $\mathcal{L}_i$ is the likelihood of event $i$ as a function of $\delta$ and $\pi(\delta)$ is the prior of the combined analysis~\cite{Mandel:2018mve,thrane2019introduction}. We do not rerun parameter estimation. We reuse the single-event posterior samples and combine them in two ways.

First is linear stacking. We pool samples from all events and assign each sample the same weight. It averages the events rather than multiplying their likelihoods, and it does not account for the different prior ranges used for each event (Table~\ref{tab:bfs}). Figure~\ref{fig:direct}
shows the distributions of $\alpha_{13}$ and $\epsilon_3$ along with their $ 99.7\%$ intervals. The median is close to zero for both parameters, and the $ 99.7\%$ range is about $-1.2$ to $1.5$ for $\alpha_{13}$ and $-4.9$ to $4.4$ for $\epsilon_3$. These ranges are no tighter than those of the best single events, since the weaker events broaden the constraints.

The second method is the pointwise product. We build a kernel density estimate (KDE) of each event's posterior, divide by that event's prior to recover its likelihood, and multiply the results on a grid of $\delta$,
\begin{equation}
  w(\delta)\;=\;\pi(\delta)\prod_{i=1}^{N}\frac{\mathrm{KDE}_i(\delta)}{\pi_i(\delta)}. \label{eq:product}
\end{equation}

\begin{figure}[h]
    \centering
	\includegraphics[width=\linewidth]{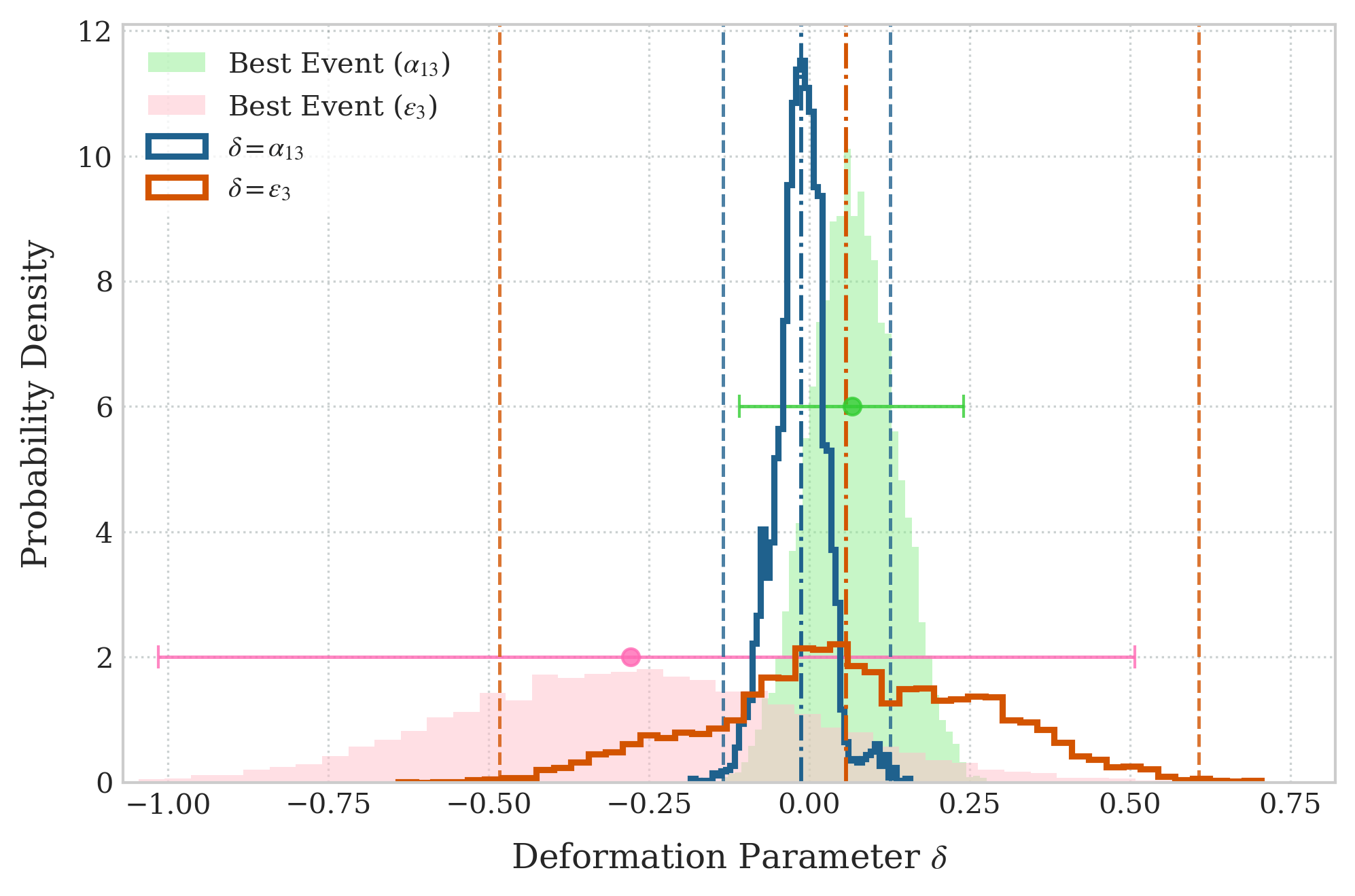} 
    \caption{Joint posterior for a universal deformation, obtained from the
pointwise product of the single-event likelihoods (Eq.~\eqref{eq:product}). The blue histogram shows $\alpha_{13}$, and the orange histogram shows $\epsilon_3$. Dot-dashed lines mark the medians and dashed lines mark the $ 99.7\%$ limits, in the corresponding colors. For comparison, the filled semi-transparent histograms show the constraints from the single-event (GW230627\_015337) analysis which yields the tightest individual bounds for both deformation parameters ($\alpha_{13}$ in green and $\epsilon_3$ in pink). The horizontal error-bar markers display the median and $ 99.7\%$ limits of this best single-event measurement in matching green and pink tones.}
    \label{fig:joint}
\end{figure}

This gives a smooth joint posterior, even where individual samples are sparse. Figure~\ref{fig:joint} shows that the constraints tighten considerably compared with stacking. At $ 99.7\%$ we find
$\alpha_{13}=-0.01^{+0.14}_{-0.12}$ and $\epsilon_3=0.06^{+0.59}_{-0.54}$. Both are consistent with zero. For comparison, Fig.~\ref{fig:joint} also shows the results from the best single event, GW230627\_015337. At 99.7\%, $\alpha_{13}\in[-0.12,\,0.24]$ (with median $\equiv 0.07$) and $\epsilon_3\in[-1.0,\,0.5]$ (median $\equiv -0.28$). The joint constraints,
$[-0.13,\,0.13]$ and $[-0.48,\,0.65]$, are only about 30\% narrower for both
parameters.

This is what we expect when one event is much more precise than the others. A product of likelihoods is dominated by its narrowest members, and
GW230627\_015337 constrains both parameters considerably better than most events. The remaining events help mainly by pulling the median, not by shrinking the interval. For $\alpha_{13}$, GW230627\_015337 prefers a small positive value ($+0.07$), while the next most precise event,
GW230605\_065343, prefers a negative one ($-0.30$). Together they help to explain why the joint median is close to zero. We see similar trend with the $\epsilon_3$ posteriors too. Also, the joint $\epsilon_3$ interval is about four times wider than the $\alpha_{13}$ interval, as it is for single event analysis, because $\epsilon_3$ leaves a weaker imprint on the phase (Eqs.~(\ref{eq:psi_a13})--(\ref{eq:psi_e3})). The modest improvement over the most constraining single event, GW230627\_015337, suggests that the joint bound is set mostly by a few precise events and not by the size of the catalog.

\subsection{Hierarchical Analysis}
\label{sec:4B}

\begin{figure*}[ht]
    \centering
        \includegraphics[width=0.49\textwidth]{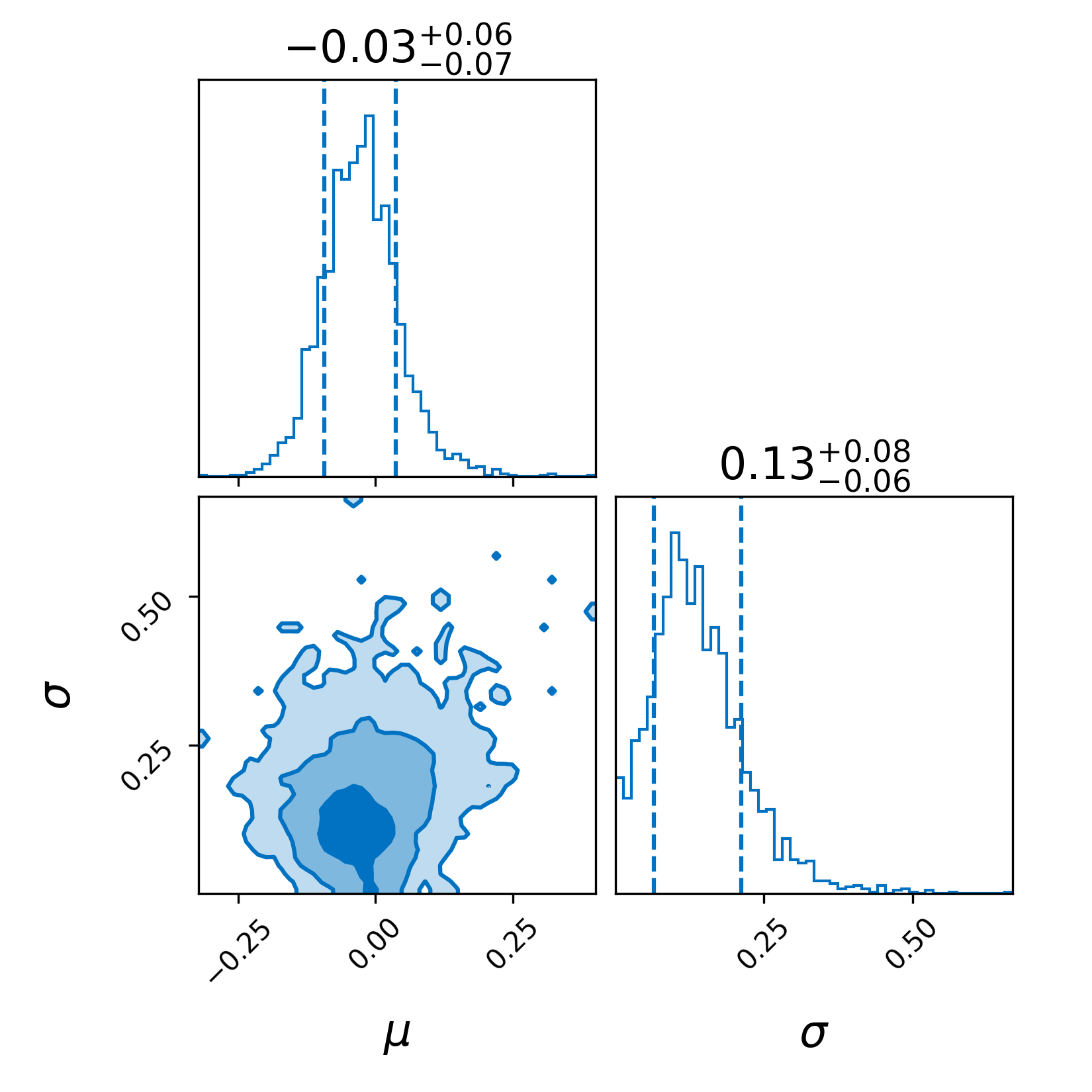}
    \hfill 
        \includegraphics[width=0.49\textwidth]{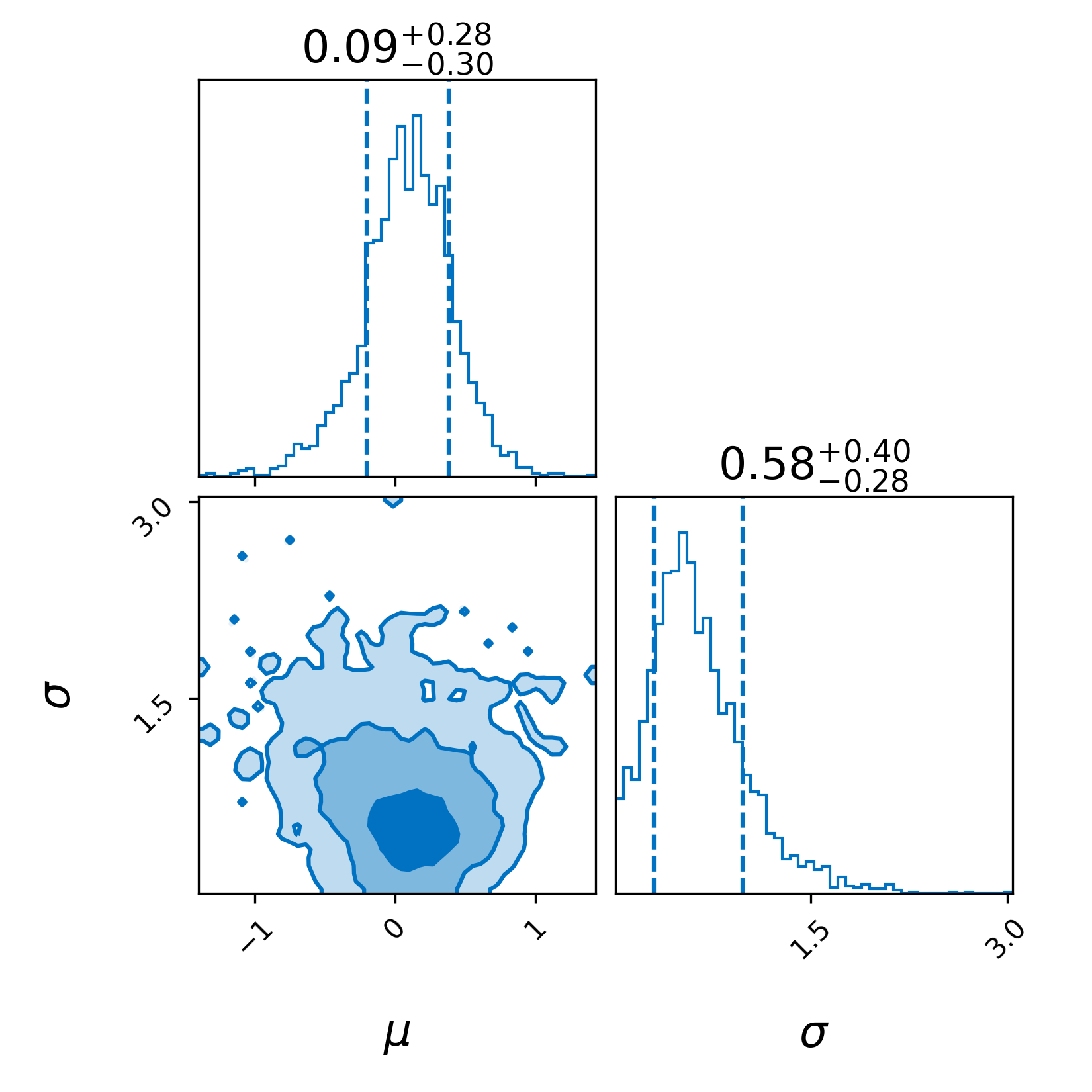}
    
    \caption{Posterior distributions of the population mean $\mu$ and scatter $\sigma$ for $\alpha_{13}$ (left) and $\epsilon_3$ (right), each obtained with the other parameter set to zero. Dashed lines in the one-dimensional panels mark the 68\% credible interval, which is also quoted above each panel.}
    \label{fig:hier}
\end{figure*}

Instead of assuming the universal nature of the deformation parameters, we now treat the true deformations as if they are drawn from a population, $\delta_i\sim\mathcal{N}(\mu,\sigma)$~\cite{Isi:2019asy,Isi:2022cii,Mancarella:2025uat}, generalizing to multidimensional hierarchical formalisms across parameters \cite{Zhong:2024pwb}. In our study, we focus on the first two moments. Here $\mu$ is the typical deformation and $\sigma$ is how much it varies between sources. We infer $(\mu,\sigma)$ from all events and judge agreement with Kerr by the posterior at $\mu=\sigma=0$. The earlier pictures are special points of this model, $\mathcal{N}(\mu,\sigma)$. Kerr case is $(0,0)$. A universal deformation is $(\mu,0)$, and pure source-to-source variation of the deformations is $(\mu,\sigma)$ with $\sigma>0$.

If each measurement $x_i$ of event $i$ is Gaussian with width $\sigma_i$, the true $\delta_i$ follows the distribution~\cite{Isi:2022cii}
\begin{equation}
\mathcal{N}\!\left(x_i;\mu,\sigma_i^2,\sigma^2\right) = \frac{1}{\sqrt{2 \pi (\sigma_i^2+\sigma^2)}} exp{\left( - \frac{(x_i-\mu)^2}{2 (\sigma_i^2+\sigma^2)} \right)}.
 \label{eq:hier_gauss}
\end{equation}

The scatter we observe consists of two components, measurement noise and intrinsic scatter. Our posteriors are not exactly Gaussian, so we work directly with the samples. For each event, we average the population Gaussian over its posterior samples, after dividing out the event's prior $\pi_i$, and then multiply over events~\cite{Mandel:2018mve,thrane2019introduction},
\begin{equation}
 \mathcal{L}(\mu,\sigma)\;\propto\;\prod_{i=1}^{N}\Big\langle \frac{\mathcal{N}(\delta;\,\mu,\,\sigma)}{\pi_i(\delta)}\Big\rangle.
 \label{eq:hier_samples}
\end{equation}

For Gaussian posteriors, this reduces to Eq.~\eqref{eq:hier_gauss}. Because the single-event priors are divided out, the result does not depend on the prior ranges that affected the Bayes factors of Sec.~\ref{sec:3}. The choice is passed to the hyperpriors, which are uniform over [-5, 5].

We implemented Eq.~\eqref{eq:hier_samples} as a custom \bilby likelihood class~\cite{Ashton:2018jfp} and sampled $(\mu,\sigma)$ with \dynesty~\cite{Speagle:2019ivv}, using $10^4$ live points. We used all 17 events, analyzing $\alpha_{13}$ first and then $\epsilon_3$, with the other parameter set to zero. We do not model selection effects because the 17 events in our study had been used in Refs.~\cite{Das:2024mjq,Das:2026zyt} based on specific selection criteria. Their impact on tests of this kind is studied in Ref.~\cite{Magee:2023muf}.

It helps to know in advance what 17 events can resolve (complementing analytical frameworks such as those in Ref.~\cite{Gair:2022fsj}). For $N$ events of similar precision $\sigma_{\rm obs}$, the uncertainties scale as~\cite{Isi:2022cii}
\begin{equation}
 \mathrm{var}(\mu)=\frac{\sigma_{\rm obs}^2+\sigma^2}{N},\qquad
 \mathrm{var}(\sigma^2)=\frac{2}{N}\left(\sigma_{\rm obs}^2+\sigma^2\right)^2 .
 \label{eq:hier_scaling}
\end{equation}

The mean improves as $1/\sqrt{N}$. The scatter is harder to measure. If $\sigma\ll\sigma_{\rm obs}$, the uncertainty on $\sigma^2$ is about $\sqrt{2/N}\,\sigma_{\rm obs}^2$, so $\sigma$ is resolved only if it exceeds roughly $(2/N)^{1/4}\sigma_{\rm obs}\approx0.6\,\sigma_{\rm obs}$ for $N=17$.

Figure~\ref{fig:hier} shows the corner plots of the hyperparameters $\mu$ and $\sigma$, with 68\% intervals. We infer $\mu_{\alpha_{13}}=-0.03^{+0.06}_{-0.07}$ and $\mu_{\epsilon_3}=0.09^{+0.28}_{-0.30}$. Both are consistent with zero, so there is no sign of a common offset. For the scatter we find $\sigma_{\alpha_{13}}=0.13^{+0.08}_{-0.06}$ and $\sigma_{\epsilon_3}=0.58^{+0.40}_{-0.28}$. In both cases, $\sigma$ peaks away from zero and $\sigma=0$ lies outside the 68\% interval. However, the posteriors keep substantial weight near zero.

We cannot claim with confidence that the scatter is real. Several things can produce it. First, the most precise events do not quite agree. GW230627\_015337 and GW230605\_065343 differ in $\alpha_{13}$ by more than $2\sigma$ (Table~1 of Ref.~\cite{Das:2026zyt}), and a nonzero $\sigma$ could be accommodating this. Secondly, noise and waveform systematics (Sec.~\ref{sec:3C}) scatter events relative to one another~\cite{LIGOScientific:2026fcf}, and the model cannot distinguish instrumental scatter from physical scatter.
 
In brief, the population is centered on Kerr. Large deformations, shared or varying, are excluded. With 17 events, we cannot yet separate a small genuine scatter from event-level noise.

\section{Discussion} \label{sec:5}

The central question we asked in our study is whether the Johannsen deformations $\alpha_{13}$ and $\epsilon_3$ are zero, have non-vanishing values shared by all black holes, or their values different from source to source. We used 17 BBHs from GWTC-3 and GWTC-4, and worked entirely from the existing single-event posteriors. Event by event, 15 (14) of the 17 events favor Kerr over the deformed model for $\alpha_{13}$ ($\epsilon_3$), but no single event does so decisively. The prior-sensitivity tests show why. A single event cannot exclude deformations of order $|\alpha_{13}|\sim0.5$, and for $\epsilon_3$ the reach is several times worse. Multiplying the events gives very large factors in favor of Kerr, but this compares Kerr with independent deformations drawn from the full prior. It measures the resolving power of the data, not the probability that Kerr is correct. No single event dominates the product. Two events, GW230605\_065343 and GW230904\_051013, and a third for $\epsilon_3$, GW230628\_231200, lean mildly towards a deformation. We found no study that explains this, and the LVK find the same behavior for GW230628\_231200 to be statistically expected among their events~\cite{LIGOScientific:2026fcf}.

If all black holes share one deformation, our set of events gives $\alpha_{13}=-0.01^{+0.14}_{-0.12}$ and $\epsilon_3=0.06^{+0.59}_{-0.54}$ at $ 99.7\%$. This is only modestly tighter than the best single event, because a few precise events dominate this estimate. In the hierarchical analysis, the population means $\mu$s are consistent with zero for both parameters. The scatter is not resolved. Its posterior peaks at 0.13 for $\alpha_{13}$ and 0.58 for $\epsilon_3$, but 17 events can resolve scatter only above about $0.6\,\sigma_{\rm obs}$, and our values sit at that edge.

Our answer to the question in the title is therefore this. The data are consistent with all black holes being Kerr black holes. They limit how large a shared deformation can be, to about 0.1 for $\alpha_{13}$ and 0.5 for $\epsilon_3$. They cannot yet separate a shared deformation from a varying one, and they show no strong evidence for either.

Several caveats apply in our mathematical treatment. We assign the same $\delta$ to both black holes in a binary, so each event constrains an effective deformation and not two separate ones. We analyze only the inspiral, with a phase correction truncated at 4PN, so waveform systematics and unmodeled spin and mass-ratio effects can matter. We vary $\alpha_{13}$ and $\epsilon_3$ one at a time, because they are nearly degenerate, so a bound on one assumes the other vanishes. The Gaussian population cannot represent a small subpopulation of deformed black holes. We do not model selection effects~\cite{Magee:2023muf}. Moreover, the combined bounds rest on a handful of events.

These results bear on a wider question. The Johannsen metric is one of several theory-agnostic parametrizations of deviations from Kerr. Others include the Johannsen--Psaltis metric~\cite{Johannsen:2011dh}, the Konoplya--Rezzolla--Zhidenko metric~\cite{Konoplya:2016jvv}, generic parametrizations of spinning geometries~\cite{Cardoso:2014rha}, and the older 'bumpy' black holes built from multipole moments~\cite{Ryan:1995wh,Vigeland:2011ji}. Reviews of such tests are given in Refs.~\cite{Yagi:2016jml,Berti:2015itd,Carson:2020rea, Cardoso:2016ryw}. Their strength lies in testing the Kerr hypothesis without committing to a particular beyond-GR theory. Their weakness is that a constraint on a parameter of one metric does not translate directly into a constraint on a theory.

The distinction we drew between a universal and a source-dependent deformation has a physical meaning. In theories with a fixed length scale $\ell$, such as higher-derivative corrections to GR~\cite{Cano:2019ore}, the deviations of a black hole of mass $M$ depend on powers of $\ell/M$. They then differ between sources, even if one theory describes every black hole. Hair that is not fixed by mass and spin~\cite{Herdeiro:2016tmi,Sotiriou:2013qea,Sotiriou:2014pfa} would also produce scatter. A universal $\delta$ is expected only when the deviation has no length scale, or when a generalized no-hair theorem holds. Seen this way, the natural next step is a population model in which the mean depends on mass or spin, $\mu=\mu(M,a)$, and not a single constant. Mixture models would let us test whether only a fraction of black holes deviate, for example those formed through particular channels~\cite{Gerosa:2021mno}.

New and future data should improve the results in two ways. The population mean improves as $1/\sqrt{N}$ with the number of events. The scatter improves more slowly, since $\sigma$ is resolvable only above $(2/N)^{1/4}\sigma_{\rm obs}$, so louder events help better than more events. GWTC-5.0, released in
May 2026, added 161 candidates from the second part of the fourth observing run (O4b), bringing the catalog to 390 events~\cite{LIGOScientific:2026wfs,LIGOScientific:2026sit}. These include the loudest signal detected so far, GW250114, with a network signal-to-noise ratio of about 80~\cite{LIGOScientific:2025rid}. The final part of the run, O4c, is planned for release in GWTC-6.0 in December 2026~\cite{LIGOScientific:2026wfs,Berry:2026qni}. Only some of these events will pass our selection criteria. A few loud events would not just help the scatter more than many weak ones, but also could break the degeneracy between $\alpha_{13}$ and $\epsilon_3$ through the $u^{-1/3}$ term, allowing both parameters to be fitted together, for which multidimensional hierarchical methods exist~\cite{Zhong:2024pwb}. Further ahead in time, third-generation detectors~\cite{ET:2019dnz,Evans:2021gyd} should deliver signals tens to hundreds of times louder, with long inspirals in the sensitive band of the detectors. Space-based detectors such as LISA~\cite{LISA:2017pwj}, TianQin~\cite{TianQin:2015yph} and Taiji~\cite{Hu:2017mde}, and multiband observations~\cite{Sesana:2016ljz,Carson:2019kkh}, would add many more inspiral cycles and would also observe supermassive black holes, whose masses are many orders of magnitude larger than those of the stellar-mass black holes in the current GWTC
catalogs~\cite{Babak:2017tow,Barausse:2020rsu,Gair:2012nm}. This would allow us to test the Scenario 3 about black holes having secondary hairs and if the deformation is dependent on the mass or spin of the black hole.

Several methodological steps would also help, such as environmental effects~\cite{Barausse:2014tra}, waveform systematics, and selection effects~\cite{Magee:2023muf} should be included in the population model. Calibrating the offsets observed in a few events by injecting GR signals into the same noise would show how often they arise by chance. Until then, the safest reading is that current GW data are consistent with the Kerr hypothesis, and that our population analysis gives a framework in which every new detection tightens that statement.

\begin{acknowledgments}
This work is supported by the National Natural Science Foundation of China (NSFC) under Grant Nos. 12347103, 12547104, and W2611007, and University of Chinese Academy of Sciences Project No. E6EQ6601X2. The work of C.B. was supported by the NSFC, Grant No.~W2531002. YH is supported by the National Key Research and Development Program of China (No. 2023YFC2206701). 
The manuscript is based upon work supported by NSF's LIGO Laboratory, which is a major facility fully funded by the National Science Foundation (NSF), as well as the Science and Technology Facilities Council (STFC) of the United Kingdom, the Max-Planck-Society (MPS), and the State of Niedersachsen/Germany for support of the construction of Advanced LIGO and construction and operation of the GEO600 detector. Additional support for Advanced LIGO was provided by the Australian Research Council. Virgo is funded through the European Gravitational Observatory (EGO), by the French Centre National de Recherche Scientifique (CNRS), the Italian Istituto Nazionale di Fisica Nucleare (INFN) and the Dutch Nikhef, with contributions by institutions from Belgium, Germany, Greece, Hungary, Ireland, Japan, Monaco, Poland, Portugal, Spain. KAGRA is supported by the Ministry of Education, Culture, Sports, Science and Technology (MEXT), Japan Society for the Promotion of Science (JSPS) in Japan; National Research Foundation (NRF) and the Ministry of Science and ICT (MSIT) in Korea; Academia Sinica (AS) and National Science and Technology Council (NSTC) in Taiwan. \\
\end{acknowledgments}

\appendix

\clearpage

\bibliography{references}

\end{document}